\documentclass[nonacm,acmlarge,authorversion]{acmart}

\usepackage[utf8]{inputenc} 
\usepackage[T1]{fontenc}    
\usepackage{url}            
\usepackage{booktabs}       
\usepackage{amsfonts}       
\usepackage{amsmath}
\usepackage{bbm}
\usepackage{nicefrac}       
\usepackage{color}
\usepackage{tikz}
\usepackage{pgfplots}
\usepackage{enumitem}

\acmBooktitle{}

\definecolor{g-blue}{HTML}{2E86C1}
\definecolor{g-red}{HTML}{B03A2E}
\definecolor{g-purple}{HTML}{AF7AC5}

\usepackage{array}
\newcolumntype{H}{>{\setbox0=\hbox\bgroup}c<{\egroup}@{}}

\newcommand\run[1]{\texttt{\small #1}}

\title{The Expando-Mono-Duo Design Pattern for Text Ranking with Pretrained Sequence-to-Sequence Models}

\author{Ronak Pradeep, Rodrigo Nogueira, and Jimmy Lin}

\affiliation{\\
David R. Cheriton School of Computer Science, University of Waterloo\\[1ex]
}

\begin{document}

\begin{abstract}
We propose a design pattern for tackling text ranking problems, dubbed ``Expando-Mono-Duo'', that has been empirically validated for a number of {\it ad hoc} retrieval tasks in different domains.
At the core, our design relies on pretrained sequence-to-sequence models within a standard multi-stage ranking architecture.
``Expando'' refers to the use of document expansion techniques to enrich keyword representations of texts prior to inverted indexing.
``Mono'' and ``Duo'' refer to components in a reranking pipeline based on a pointwise model and a pairwise model that rerank initial candidates retrieved using keyword search.
We present experimental results from the MS MARCO passage and document ranking tasks, the TREC 2020 Deep Learning Track, and the TREC-COVID challenge that validate our design.
In all these tasks, we achieve effectiveness that is at or near the state of the art, in some cases using a zero-shot approach that does not exploit any training data from the target task.
To support replicability, implementations of our design pattern are open-sourced in the Pyserini IR toolkit and PyGaggle neural reranking library.
\end{abstract}

\maketitle

\section{Introduction}

For text ranking tasks (specifically, {\it ad hoc} retrieval), a simple two-stage retrieve-then-rerank architecture has proven to be an effective and widely adopted approach~\cite{Asadi_Lin_SIGIR2013,Pedersen_SIGIR2010}.
Retrieval can be accomplished via keyword search, e.g., ranking with BM25~\cite{robertson1995okapi}, or more recently, via approximate nearest-neighbor search on learned dense representations~\cite{lee-etal-2019-latent,Karpukhin:2004.04906:2020,Gao:2004.13969:2020,Xiong:2007.00808:2020,ColBERT,Lin_etal_arXiv2020_DenseRanking}.
Reranking is typically accomplished using pretrained transformers such as BERT~\cite{devlin-etal-2019-bert} or one of its variants that have been fine-tuned with (query, relevant document) pairs~\cite{Nogueira:1901.04085:2019}.

We present a refinement of this general approach that has been empirically demonstrated to work well for multiple {\it ad hoc} retrieval tasks in different domains.
In contrast to most current approaches that build on encoder-only pretrained transformers such as BERT~\cite{devlin-etal-2019-bert}, our approach instead relies on pretrained sequence-to-sequence transformers within a multi-stage ranking architecture.
In our case, we use T5~\cite{raffel2019exploring}, but our approach can be extended to other sequence-to-sequence models such as BART~\cite{lewis-etal-2020-bart} and Pegasus~\cite{zhang2019pegasus} as well.
The key features of our approach are as follows:

\begin{itemize}

\item Document expansion using a sequence-to-sequence model to enrich keyword representations of texts from the corpus prior to indexing (``Expando''); we've also previously called this approach ``doc2query''.

\item Initial keyword-based retrieval (also called first-stage retrieval or candidate generation) using standard inverted indexes.

\item A two-stage reranking pipeline comprising a pointwise reranker (``Mono'') followed by a pairwise reranker (``Duo''), both built on pretrained sequence-to-sequence models.

\end{itemize}

\noindent This combination, which we dub ``Expando-Mono-Duo'' has been empirically validated on a wide range of {\it ad hoc} retrieval tasks in different domains.
Based on formal evaluations, our approach has achieved effectiveness at or near the state of the art, sometimes in a completely zero-shot manner (i.e., without fine-tuning models on data from the target task).
The generality of this approach, we believe, suggests that elevating it to a ``design pattern'' for text ranking might be justified.

In this paper, we provide details about each aspect of the ``Expando-Mono-Duo'' design pattern, how the design is specifically instantiated for different tasks, and report experimental results on five benchmark datasets:\ the MS MARCO passage and document ranking tasks, the passage and document ranking tasks at the TREC 2020 Deep Learning Track,  and the TREC-COVID challenge.
While some components in this pattern have been described separately and in a piece-wise manner, this is the first paper where we have brought together all these separate threads of work, which allows us to thoroughly describe our ideas in a coherent, self-contained manner and to present ablation analyses that quantify the impact of each component.

\section{Background and Related Work}

We assume the standard definition of {\it ad hoc} retrieval, where given a corpus of texts $\mathcal{C}$, the goal of a ranking model is to return a top $k$ ranked list of texts from the corpus in response to an information need $q$ that maximizes some metric of ranking quality such as nDCG or MRR.
In this paper we use the terms {\it ad hoc} retrieval and ranking interchangeably.

The basic idea behind multi-stage ranking architectures is to break {\it ad hoc} retrieval down into a series of pipeline stages.
Following an initial retrieval stage (also called candidate generation or first-stage retrieval), where a bag-of-words query is typically issued against an inverted index, each subsequent stage reranks the list of candidates passed along from the previous stage until the final top $k$ results are generated for consumption, e.g., returned to the user.
Recognizing that the unit of retrieval might differ based on the task, per standard parlance in IR, we use {\it document} to refer to the text being retrieved in a generic sense, when in actuality it may be a passage (in the case of MS MARCO passage ranking) or some hybrid construction (in the case of our TREC-COVID experiments).

Multi-stage ranking architectures have received much interest in academia~\cite{Matveeva_etal_SIGIR2006,Wang_etal_SIGIR2011,Asadi_Lin_SIGIR2013,ChenRuey-Cheng_etal_SIGIR2017a,Mackenzie_etal_WSDM2018} as well as industry.
Documented production deployments of this architecture include the Bing web search engine~\cite{Pedersen_SIGIR2010} as well as Alibaba's e-commerce search engine~\cite{LiuShichen_etal_SIGKDD2017}.
These represent instances of a mature and well-studied design, which originally evolved to strike a balance between model complexity and search latency by controlling the size of the candidate set at each stage.
Increasingly richer (and typically slower) models can be made {\it practical} by considering successively smaller candidate sets.
For certain (easy) queries, stages of the pipeline can be skipped entirely, known as ``early exits''~\cite{Cambazoglu_etal_WSDM2010}.
Viewed in this manner, multi-stage ranking captures the same intuition as progressive refinement in classifier cascades~\cite{Viola_Jones_208}, which has a long history.
For example, an early stage might consider only term statistics of individual terms, whereas later stages might consider bigrams, phrases, or even apply lightweight NLP techniques (which can better capture aspects of relevance but are more computationally intensive).
Given this setup, a number of researchers have proposed techniques based, for example, on boosting for composing these stages in an end-to-end manner~\cite{Wang_etal_SIGIR2011,XuZhixiang_etal_ICML2012}.
While these techniques were previously explored mostly in the context of feature-based learning to rank~\cite{LiuTY_FnTIR2009,LiHang_2011}, many of the same ideas remain applicable in a transformer-based setting~\cite{Nogueira_etal_arXiv2019_multistageBERT,soldaini-moschitti-2020-cascade,Xin_etal_SustaiNLP2020}.

In the past decade or so, the advent of deep learning has brought tremendous excitement to the information retrieval community.
Although machine-learned ranking models have been well studied since the mid-2000s in the context of learning to rank~\cite{LiuTY_FnTIR2009,LiHang_2011}, the paradigm was heavily driven by manual feature engineering; commercial web search engines are known to incorporate thousands of features (or more) in their models.
Continuous vector space representations coupled with neural models promised to obviate the need for handcrafted features and have attracted much attention from both researchers and practitioners.
Well-known early neural ranking models include DRMM~\citep{guo2016deep}, DUET~\citep{mitra2017learning}, KNRM~\citep{xiong2017end}, and Co-PACRR~\citep{hui2018co}; the literature is too vast for an exhaustive review here, and thus we refer readers to recent overviews~\cite{Onal_etal_IRJ2018,MitraBhaskar_Craswell_2019}.

The introduction of BERT~\cite{devlin-etal-2019-bert}, however, marked the beginning of a new era in neural ranking models, beginning with~\citet{Nogueira:1901.04085:2019}.
Over the past couple of years, we have witnessed the increasing dominance of reranking models based on pretrained transformers~\cite{Dai_Callan_SIGIR2019,macavaney2019cedr,yilmaz2019cross,Li:2008.09093:2020}.
While retrieval using approximate nearest-neighbor search on learned dense representations has emerged as a promising direction~\cite{lee-etal-2019-latent,Karpukhin:2004.04906:2020,Gao:2004.13969:2020,Xiong:2007.00808:2020,ColBERT,Lin_etal_arXiv2020_DenseRanking}, a multi-stage ranking architecture based on keyword search as first-stage retrieval remains popular.
For more details, the recent survey by~\citet{Lin_etal_arXiv2020_ptr4tr} provides an overview of these developments.

Although this paper represents the first opportunity we have had to detail the ``Expando-Mono-Duo'' design pattern, individual components have been presented elsewhere, albeit in a piece-wise manner.
We were the first to propose multi-stage ranking with transformer models~\cite{Nogueira_etal_arXiv2019_multistageBERT}, although in the context of BERT (contrasting with our shift to sequence-to-sequence models here) and without document expansion.
The ``Expando'' idea originated in~\citet{doc2query-base} and was subsequently improved on in~\citet{docTTTTTquery}.
``Mono'' is detailed in~\citet{Nogueira_etal_FindingsEMNLP2020}.
``Duo'' with BERT appears in~\citet{Nogueira_etal_arXiv2019_multistageBERT}, and was also referenced in a description of our submissions to TREC-COVID~\cite{ZhangEdwin_etal_SDP2020}.
However, none of these previous papers contained a clear end-to-end explication of our ideas, presentation of comprehension evaluation results, and ablations analyzing the contributions of different components.

\section{Expando-Mono-Duo with T5}

In our formulation, a multi-stage ranking architecture comprises a number of stages, which we denote $H_0$ to $H_N$.
Except for $H_0$, which retrieves $k_0$ candidates based on keyword search (i.e., from an inverted index), each stage $H_n$ receives a ranked list $R_{n-1}$ comprising $k_{n-1}$ candidates from the previous stage.
Each stage, in turn, provides a ranked list $R_n$ comprising $k_n$ candidates to the subsequent stage, with the obvious requirement that $k_n \le k_{n-1}$.
The ranked list generated by the last stage $H_N$ in the pipeline is designated for final consumption, i.e., shown to the human searcher or fed to a downstream application.
In this context, $k_N$ is the desired $k$ in the original ranking task; all the other $k_n$'s $n \in \{0 \ldots N-1\}$ are internal system settings.
As a simple example, for the MS MARCO passage ranking task, where the official metric is MRR@10 (that is, the evaluation metric only considers the first ten hits), a popular setting is to retrieve 1000 hits using keyword search, which are then fed to a reranker to produce the top 10 final results~\cite{Nogueira:1901.04085:2019}.

\begin{figure*}[t]
\begin{center}
\centerline{\includegraphics[width=0.8\textwidth]{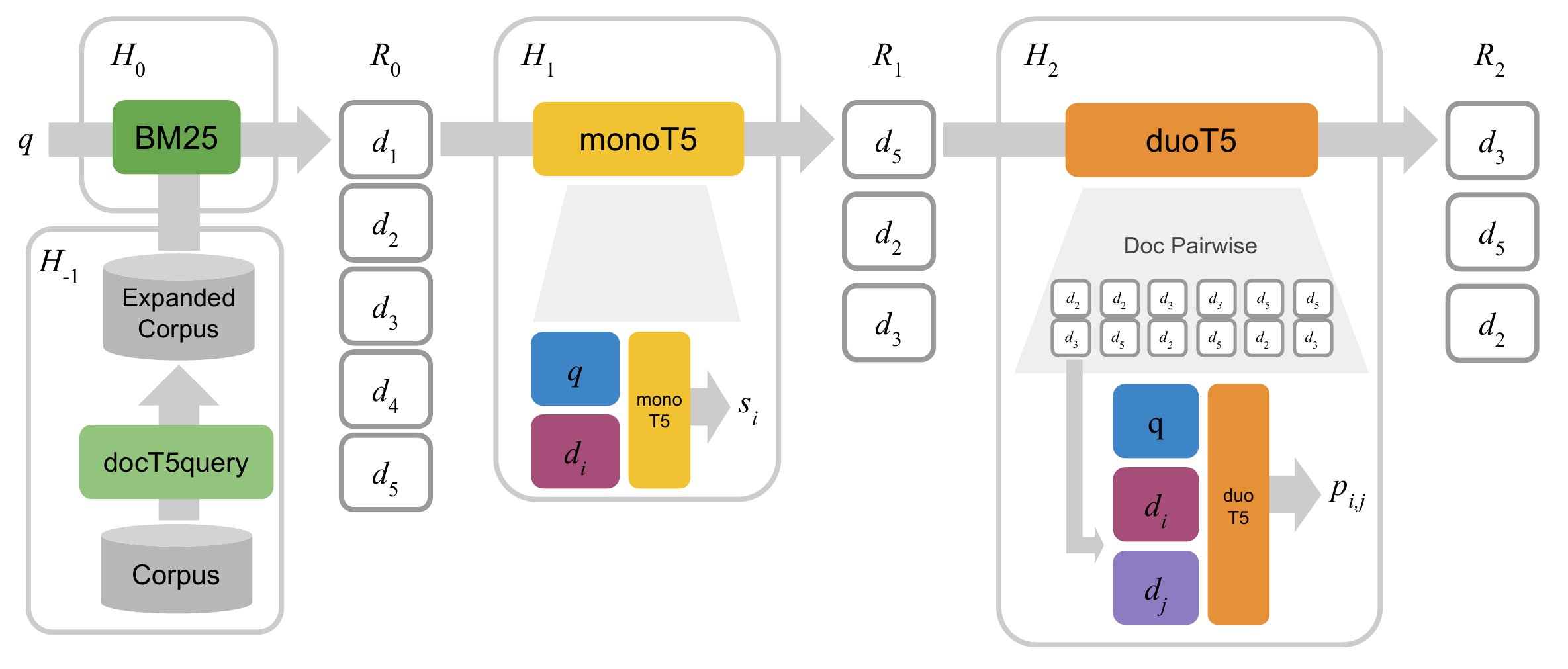}}
\caption{Illustration of our multi-stage ranking architecture. Prior to indexing, we perform document expansion, denoted $H_{-1}$. In stage $H_0$, given a query $q$, the top $k_0$ ($=5$ in the figure) candidate documents $R_0$ are retrieved using BM25. In stage $H_1$, monoT5 produces a relevance score $s_i$ for each pair of query $q$ and candidate $d_i \in R_0$. The top $k_1$ ($=3$ in the figure) candidates with respect to these relevance scores are passed to stage $H_2$, in which duoT5 computes a relevance score $p_{i,j}$ for each triple ($q$, $d_i$, $d_j$). The final output $R_2$ is formed by reranking the candidates according to these scores (see Section~\ref{section:duoT5} for how these pairwise scores are aggregated).} 
\label{fig:overview}
\end{center}
\end{figure*}

Prior to building the inverted index that feeds first-stage retrieval $H_0$, with ``Expando'', we first perform document expansion on the input corpus to enrich its representation; we denote this as the $H_{-1}$ stage.
The augmented documents are then indexed exactly as before; the only impact of document expansion is to (hopefully) provide a richer set of candidate documents for downstream rerankers to process.
The output of first-stage retrieval $H_0$ is then passed to the two-stage reranking pipeline comprised of monoT5 $H_1$ (``Mono'') and duoT5 $H_2$ (``Duo'').
We describe each component of the overall architecture (see Figure~\ref{fig:overview}) in detail below.
Here, our narrative focuses on the overall design, and we defer details of our experimental settings for each task to Section~\ref{sec:setup}.
To support replicability, all our implementations are open-sourced in the Pyserini IR toolkit\footnote{\url{http://pyserini.io/}} and PyGaggle\footnote{\url{http://pygaggle.ai/}} neural reranking library.

\subsection{$H_{-1}$: Document Expansion with T5}

The idea behind document expansion is to enrich each document (more generally, each text from the corpus) with additional text (containing keywords) that are representative of its content for the purposes of retrieval.
In our particular implementation, we leverage a corpus of (query, relevant passage) pairs to train a sequence-to-sequence model that maps passages to queries.
That is, given a segment of text, the model predicts queries that the text can potentially answer.
For example, consider the following passage:

\begin{quote}
July is the hottest month in Washington DC with an 
average temperature of 27$^{\circ}$C (80$^{\circ}$F) and the coldest is
January at 4$^{\circ}$C (38$^{\circ}$F) with the most daily sunshine hours at 9
in July. The wettest month is May with an average of 100mm of rain.
\end{quote}

\noindent In this example, the ``target query'' (i.e., from our training data) is ``What is the temperature in Washington?''
In the MS MARCO dataset we used for training, the queries tend to be relatively well-formed natural language utterances, but for generality we refer to model output as queries.
After learning from a large corpus of such examples (details later), our model is able to predict the query ``What is the weather in Washington DC?''

These predictions are then directly appended to the original (source) document; once this expansion has been performed for every document, the new augmented corpus is indexed, as before.
Since document expansion occurs before indexing, we refer to this as the $H_{-1}$ stage.
Note that in this case, the predicted query contains a term (``weather'') that is not present in the original passage.
This has the effect of enriching the passage and increasing the likelihood of matching a broader range of queries for which this passage would be relevant.
Even in the case where the predicted queries repeat words that are already present in the text, it achieves the effect of increasing the weight on the term, highlighting its importance.

This idea, dubbed ``doc2query'', was first proposed in~\citet{doc2query-base}, but in this work we take advantage of the pretrained sequence-to-sequence model T5, as described in~\citet{docTTTTTquery}.
In that paper, it was given the somewhat awkward name docTTTTTquery (also written as docT5query), which seemed like a whimsical yet accurate description at the time, although in retrospect the moniker is unwieldy in prose.
Here, we refer to the document expansion model with T5 as doc2query-T5, to disambiguate it from the original doc2query proposal~\cite{doc2query-base}, which used vanilla (non-pretrained) transformers.

\subsection{$H_0$: Keyword Retrieval}

The stage $H_0$ receives as input the user query $q$ and produces top $k_0$ candidates $R_0$.
In multi-stage ranking architectures this is called first-stage retrieval or candidate generation.
In our implementation, the query is treated as a bag of words for ranking documents from the corpus using a standard inverted index based on BM25~\cite{robertson1995okapi}.
Depending on the exact setting and task (see Section~\ref{sec:setup}), keyword search may additionally exploit query expansion using pseudo-relevance feedback.
All our experiments used the Pyserini IR toolkit,\footnote{\url{http://pyserini.io/}} which is the Python wrapper to Anserini~\cite{Yang_etal_SIGIR2017,Yang_etal_JDIQ2018},\footnote{\url{http://anserini.io/}} itself built on the popular open-source Lucene search engine with the goals of supporting replicable IR research and bringing research practices into better alignment with real-world search applications.

\subsection{$H_1$: Pointwise Reranking with monoT5}

In stage $H_1$, documents retrieved in $H_0$ are reranked by a pointwise reranker called monoT5.
The model estimates a score $s_i$ quantifying how relevant a candidate $d_i \in R_{n-1}$ is to a query $q$.
That is:
\begin{equation}
P(\textrm{Relevant}=1 | d_i, q).
\end{equation}

\noindent This approach is called relevance classification~\cite{Lin_etal_arXiv2020_ptr4tr}, or alternatively, a pointwise approach in the parlance of learning to rank~\cite{LiuTY_FnTIR2009,LiHang_2011}.
Naturally, we expect that the ranking induced by these scores yields a higher metric (e.g., MAP or MRR) than the scores from the input ranking (i.e., the output of $H_0$).
At a high level, monoT5 is a sequence-to-sequence adaptation of the monoBERT model proposed by~\citet{Nogueira:1901.04085:2019} and further detailed in~\citet{Lin_etal_arXiv2020_ptr4tr}.

Details of monoT5 are described in~\citet{Nogueira_etal_FindingsEMNLP2020}; here, we only provide a short overview.
Unlike monoBERT~\cite{Nogueira:1901.04085:2019}, monoT5 uses T5~\cite{raffel2019exploring}, a popular pretrained sequence-to-sequence transformer model.
In this model, all target tasks are cast as sequence-to-sequence tasks.
Specifically, ranking is performed using the following input sequence template:
\begin{equation}
\text{Query: } q \ \ \text{ Document: } d \ \ \text{ Relevant:}
\end{equation}

\noindent where $q$ and $d$ are the query and document texts, respectively.
The model is fine-tuned to produce the tokens ``true'' or ``false'' depending on whether the document is relevant or not to the query.
That is, ``true'' and ``false'' are the ``target tokens'' (i.e., ground truth predictions in the sequence-to-sequence transformation).

At inference time, to compute probabilities for each query--document pair (in a reranking setting), we apply a softmax only on the logits of the ``true'' and ``false'' tokens.
Following~\citet{Nogueira_etal_FindingsEMNLP2020}, we rerank the documents according to the probabilities assigned to the ``true'' token.
Note that while $H_0$ uses a corpus enriched by document expansion, documents in $R_0$ consist of original texts that do not include the predicted queries.

As discussed in~\citet{Lin_etal_arXiv2020_ptr4tr}, one reoccurring theme in the application of transformers to text ranking is the handling of texts that are longer than the input sequences that the models were designed to handle (typically, 512 tokens).
Building on previous work~\cite{Hearst_SIGIR1993,Callan_SIGIR1994,Dai_Callan_SIGIR2019}, we were able to devise simple yet effective solutions to address this issue.
Since these solutions are corpus dependent, we save detailed discussion for Section~\ref{sec:setup}.

\subsection{$H_2$: Pairwise Reranking with duoT5}
\label{section:duoT5}

The output $R_1$ from the previous stage is used as input to the pairwise reranker we call duoT5.
In this pairwise approach, the reranker considers a pair of documents $(d_i, d_j)$ and estimates the probability $p_{i,j}$ that candidate $d_i$ is more relevant than $d_j$ to query $q$:
\begin{equation}
P( d_i \succ d_j | d_i, d_j, q),
\end{equation}

\noindent where $d_i \succ d_j$ is a commonly adopted notation in IR for stating that $d_i$ is {\it more relevant} than $d_j$ (with respect to the query $q$).

The basic idea behind duoT5 was originally developed in~\citet{Nogueira_etal_arXiv2019_multistageBERT}, but in the context of BERT (not surprisingly, called duoBERT).
As the name suggests, duoT5 is also based on T5.
In this case, the reranker takes as input the sequence:
\begin{equation}
\label{eq:duo_format}
\text{Query: } q \ \ \text{ Document0: } d_i
\text{ Document1: } d_j \ \text{ Relevant:}
\end{equation}
The pairwise sequence-to-sequence model is fine-tuned to produce the token ``true'' if document $d_i$ is more relevant than $d_j$, and ``false'' otherwise.

At inference time, we aggregate the pairwise scores $p_{i,j}$ so that each document receives a single score $s_i$. 
We investigated four different aggregation techniques:

\begin{equation}
\label{eq:pairwise_sum}
\textsc{Sum}: s_i = \sum_{j \in J_i} p_{i,j},
\end{equation}

\begin{equation}
\label{eq:pairwise_log_sum}
\textsc{Sum-log}: s_i = \sum_{j \in J_i} \log{p_{i,j}},
\end{equation}

\begin{equation}
\label{eq:pairwise_sym_sum}
\textsc{Sym-Sum}: s_i = \sum_{j \in J_i} \left( p_{i,j} + (1-p_{j,i}) \right),
\end{equation}

\begin{equation}
\label{eq:pairwise_sym_sum_log}
\textsc{Sym-Sum-Log}: s_i = \sum_{j \in J_i} \left( \log{p_{i,j}} + \log{(1-p_{j,i})} \right),
\end{equation}
\noindent where $J_i=\{0 \leq j < k_1, j \neq i\}$.

The \textsc{Sum} method measures the pairwise agreement that candidate $d_i$ is more relevant than the rest of the candidates ${\{d_j\}}_{j \neq i}$.
The $\textsc{Sum-Log}$ method is a variant of \textsc{Sum} that penalizes smaller values of $p_{i,j}$ by virtue of the logarithm.
The methods \textsc{Sym-Sum} and $\textsc{Sym-Sum-Log}$ are variants of \textsc{Sum} and $\textsc{Sum-Log}$, respectively.
In these two cases, we also add the score corresponding to the probability $p_{j,i}$ assigned to the ``false'' token.
The goal of these variants is to ensure a more stable and accurate scoring.
We empirically compared these variants and examine the impact of different parameter settings in Section~\ref{sec:marco_passage_results}.

The candidates in $R_1$ (i.e., the output from monoT5 stage $H_1$) are reranked according to their scores ${s_i}$ to obtain the final list of candidates $R_2$.
Note that duoT5 is computationally expensive, where the running time is dominated by the need to perform inference $k_1 \times (k_1 - 1)$ times (in comparison, the cost of computing aggregations is minimal).
Thus, in practice it is tractable to run duoT5 reranking only on a relatively small number of candidates $k_1$ (detailed settings are presented in Section~\ref{sec:setup}).
The implication of this design is that duoT5 is primarily aimed at improving early precision, i.e., the quality of results high in the ranked list.

While in principle multi-stage ranking architectures can have an arbitrary number of stages, in our current design the output $R_2$ is intended for final consumption, e.g., by a human searcher or fed to a downstream application.
In our evaluations, we report metrics computed over the output of duoT5.

\section{Experimental Settings}
\label{sec:setup}

We have empirically validated the Expando-Mono-Duo design pattern in five different settings:\ the MS MARCO passage and document ranking tasks, the passage and document ranking tasks in the TREC 2020 Deep Learning Track, and the TREC-COVID challenge.
In this section, we describe detailed experimental settings for each task, which correspond to how our Expando-Mono-Duo pattern is ``instantiated'' in different contexts.

\subsection{MS MARCO Passage Ranking}
\label{sec:setup:passage}

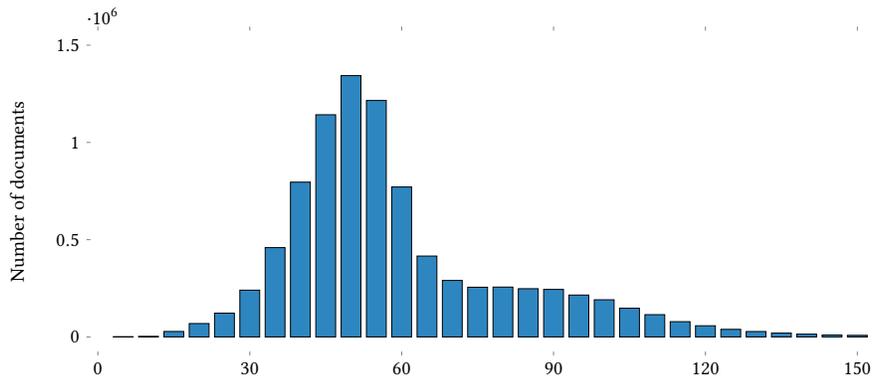
\begin{figure*}[t]
\begin{center}
\begin{tikzpicture}[scale=0.75]
  \begin{axis}[
    ybar,
    bar width=10pt,
    width=\textwidth,
    height=\axisdefaultheight,
    ylabel=Number of documents,
    y axis line style = { opacity = 0 },
    ymin=0, ymax=1500000,
    xtick={0,30,...,150},
    tickwidth         = 2pt,
    enlarge y limits  = 0.05,
    enlarge x limits  = 0.05,
    ticklabel style={
        /pgf/number format/fixed,
        /pgf/number format/precision=5
    }, 
    scaled ticks=true
  ]
  \addplot [fill=g-blue] coordinates { 
        (5, 1092)
        (10, 2819)
        (15, 28192)
        (20, 68908)
        (25, 122343)
        (30, 240444)
        (35, 459230)
        (40, 795835)
        (45, 1142273)
        (50, 1343464)
        (55, 1215627)
        (60, 771052)
        (65, 415882)
        (70, 290450)
        (75, 255577)
        (80, 256427)
        (85, 247988)
        (90, 244424)
        (95, 215443)
        (100, 191125)
        (105, 147667)
        (110, 114596)
        (115, 78525)
        (120, 57549)
        (125, 39036)
        (130, 27905)
        (135, 19947)
        (140, 14396)
        (145, 10084)
        (150, 8294)
  };
  \end{axis}
\end{tikzpicture}
\caption{Distribution of passages lengths in the MS MARCO passage corpus.}
\label{figure:ms-marco-passage:corpus_length_distribution}
\end{center}
\end{figure*}

We trained and evaluated our models on the MS MARCO passage ranking dataset~\cite{nguyen2016ms}, which provides a corpus comprised of 8.8M passages gathered from Bing search engine results.
The mean length of each passage is 56 tokens (median: 50, max: 362), and the distribution of passage lengths is shown in Figure~\ref{figure:ms-marco-passage:corpus_length_distribution}.\footnote{Tokenization here is performed using Python's string \texttt{split} method. Note that T5 uses different tokenization, and therefore these lengths do not correspond to model input lengths.} 
The training set contains approximately 532.7K (query, relevant passage) pairs, where each query has one relevant passage on average.
The development set contains 6980 queries and the test set contain 6837 queries, but relevance labels are only publicly available for the development set; evaluation on the held-out test set requires submission to the leaderboard.\footnote{\url{http://www.msmarco.org/}}

We also evaluated our models on the passage ranking task of the TREC 2020 Deep Learning Track~\cite{TREC2020_DL_overview}, which has 54 queries with graded relevance judgements (on a four-point scale) gathered via traditional pooling techniques~\cite{Voorhees_CLEF2002} (with $\sim$210 judgments per query).
The TREC evaluation used the same corpus as the MS MARCO passage ranking task, and thus the two evaluations differed primarily in the number of queries and the abundance of relevance judgments per query.
The MS MARCO relevance judgments are sometimes referred to as ``sparse'' judgments, contrasting with the richer ``dense'' judgments gathered via pooling by TREC assessors.
We adopt this terminology in our discussions.

Note that in our experiments the same exact settings were used for both the MS MARCO passage ranking task as well as the TREC 2020 Deep Learning Track.
Specifically, we did not take advantage of relevance judgments from the TREC 2019 Deep Learning Track for either evaluation.

\medskip \noindent {\bf ``Expando'' and Keyword Retrieval Settings.}
In the first stage of our pipeline, we expanded all documents in the MS MARCO passage corpus with queries predicted by our doc2query-T5 model~\cite{docTTTTTquery}, which was prepared as follows:
Starting with a publicly available checkpoint, we fine-tuned T5 with (query, relevant passage) pairs from the MS MARCO passage ranking dataset with a constant learning rate of $10^{-3}$ for 4K iterations with batches of 256, which corresponds to two epochs with the training data.
We used a maximum of 512 input tokens and 64 output tokens.
In the MS MARCO dataset, none of the inputs or outputs had to be truncated using these length settings (see Figure~\ref{figure:ms-marco-passage:corpus_length_distribution}).
Similar to~\citet{docTTTTTquery}, we found that the top-$k$ sampling decoder~\cite{fan2018hierarchical} produced better queries (i.e., leading to higher ranking effectiveness) than beam search.
We used $k = 10$ and sampled 40 queries per document with T5-base; the large model variant did not appear to yield any improvements in retrieval effectiveness.
Our online replication guide details all these settings.\footnote{\url{http://doc2query.ai/}}
We did not experiment with T5-3B and T5-11B due to their size and associated computational cost.

We used a Google TPU v3-8 to fine-tune the model and perform inference.
Training took less than 1.5 hours on a single TPU.
For inference, sampling five queries for each of the 8.8M passages in the corpus took approximately 40 hours on a single TPU.
Note that inference is trivially parallelizable and the inference time is linear with respect to the number of samples.

All expanded texts were then indexed with the Anserini IR toolkit.
The predicted queries were appended to the original passages, without any intervening delimiters.
Retrieval was performed using BM25 with the parameters $k_1=0.82$, $b=0.68$, based on tuning on the development set via simple grid search.
For the TREC 2020 Deep Learning Track topics, we additionally applied the BM25 + RM3 model for query expansion using pseudo-relevance feedback (with all default parameters in Anserini).
This approach was described in~\citet{Yang_etal_SIGIR2019} and has been shown to be a strong baseline, especially compared to pre-BERT neural ranking models, when evaluated on traditional TREC topics with ``dense'' judgements.

\medskip \noindent {\bf ``Mono'' Settings.}
As part of the MS MARCO passage dataset, the creators provided (query, relevant passage, non-relevant passage) triples, where the negative evidence came from sampling.
For training monoT5, we used the ``small'' triples file containing 39.8M records.
We fine-tuned the base, large, and 3B variants of our monoT5 models with a constant learning rate of $10^{-3}$ for 100K iterations with class-balanced batches of size 128.
We used a maximum of 512 input tokens and two output tokens (one for the true/false token and another for the end-of-sequence token); none of the inputs had to be truncated when using this length (see Figure~\ref{figure:ms-marco-passage:corpus_length_distribution}).
Training the base, large, and 3B models took approximately 12, 48, and 160 hours on a single Google TPU v3-8, respectively.

At reranking time, our model considered the top 1000 hits from first-stage retrieval (i.e., $k_0 = 1000$).
This is a commonly used setting first proposed in~\citet{Nogueira:1901.04085:2019}.
Note that the input to our reranker, here and in all our experiments, did not include the predicted (expansion) queries; that is, the goal of document expansion is to improve first-stage retrieval only.
There are two reasons for this design decision:
First, the predicted queries are sometimes noisy and can potentially mislead the model.
Second, they add to the length of each passage, thus overflowing the model input length limitations for inference.

\medskip \noindent {\bf ``Duo'' Settings.}
We fine-tuned duoT5 using the monoT5 model trained on the MS MARCO passage data as a starting checkpoint, using the same hyperparameters as those used for training monoT5.
In more detail, we also used the ``small'' triples file provided by the MS MARCO organizers, converted into duoT5's input format (Equation~\ref{eq:duo_format}) and fed to the model.
Based on initial experiments with duoT5-base, we found that model effectiveness converged around 50K iterations.
Hence, we trained duoT5-3B for 50K iterations, which took about 80 hours overall on a single Google TPU v3-8.

On the development set, we examined different numbers of candidates $k_1$ (i.e., output from the mono stage) that are reranked by the pairwise ranker.
Note that the duo model is computationally expensive, requiring a number of inferences per query that is quadratic with respect to the number of candidate pairs.
Thus, for tractability, we only ran experiments where $k_1$ ranged from 10 to 50, in increments of 10. 
The results of these explorations are described later (see Figure~\ref{fig:duot5_cutoff}); in our final experiments, we used $k_1 = 50$ with $\textsc{Sym-Sum}$ as the pairwise aggregation method.
Note that if more than $k_1$ hits are requested as the output of duoT5, we simply take additional ranked output from monoT5.
For example, if the user (or downstream application) requires 1000 hits, then the first 50 will come from duoT5 (assuming $k_1 = 50$), while the remaining results (rank positions 51--1000)\ will be the unaltered rankings from monoT5.
Thus, as explained in Section~\ref{section:duoT5}, our pairwise reranker emphasizes early precision.

\subsection{MS MARCO Document Ranking}
\label{sec:marco_doc_setup}

\begin{figure*}[t]
\begin{center}
\begin{tikzpicture}[scale=0.75]
  \begin{axis}[
    ybar,
    bar width=2.0pt,
    width=\textwidth,
    height=\axisdefaultheight,
    ylabel=Number of documents,
    xlabel=Number of tokens,
    y axis line style = { opacity = 0 },
    ymin=0, ymax=160000,
    xtick={0,1000,...,5000},
    tickwidth         = 2pt,
    enlarge y limits  = 0.05,
    enlarge x limits  = 0.05,
    ticklabel style={
        /pgf/number format/fixed,
        /pgf/number format/precision=5
    }, 
    scaled ticks=true
  ]
  \addplot [fill=g-blue] coordinates { 
        (50, 83770)
        (100, 119060)
        (150, 153865)
        (200, 154430)
        (250, 152887)
        (300, 153622)
        (350, 149536)
        (400, 145752)
        (450, 142087)
        (500, 134525)
        (550, 132263)
        (600, 121911)
        (650, 112380)
        (700, 101264)
        (750, 92831)
        (800, 82863)
        (850, 74954)
        (900, 66448)
        (950, 60933)
        (1000, 55492)
        (1050, 51626)
        (1100, 47117)
        (1150, 42815)
        (1200, 39188)
        (1250, 35771)
        (1300, 33550)
        (1350, 30908)
        (1400, 28501)
        (1450, 25937)
        (1500, 24320)
        (1550, 22830)
        (1600, 20989)
        (1650, 20084)
        (1700, 18540)
        (1750, 17444)
        (1800, 16637)
        (1850, 15691)
        (1900, 14878)
        (1950, 14367)
        (2000, 13568)
        (2050, 12647)
        (2100, 12102)
        (2150, 11607)
        (2200, 10928)
        (2250, 10456)
        (2300, 9698)
        (2350, 9032)
        (2400, 8723)
        (2450, 8516)
        (2500, 8194)
        (2550, 7420)
        (2600, 7127)
        (2650, 6820)
        (2700, 6607)
        (2750, 6322)
        (2800, 6135)
        (2850, 5805)
        (2900, 5658)
        (2950, 5313)
        (3000, 5145)
        (3050, 5097)
        (3100, 4839)
        (3150, 4824)
        (3200, 4650)
        (3250, 4373)
        (3300, 4206)
        (3350, 4072)
        (3400, 4005)
        (3450, 3810)
        (3500, 3544)
        (3550, 3634)
        (3600, 3428)
        (3650, 3385)
        (3700, 3262)
        (3750, 3084)
        (3800, 3039)
        (3850, 2910)
        (3900, 2899)
        (3950, 2829)
        (4000, 2647)
        (4050, 2655)
        (4100, 2657)
        (4150, 2417)
        (4200, 2374)
        (4250, 2368)
        (4300, 2383)
        (4350, 2274)
        (4400, 2261)
        (4450, 2075)
        (4500, 2162)
        (4550, 2007)
        (4600, 2039)
        (4650, 1913)
        (4700, 1878)
        (4750, 1801)
        (4800, 1778)
        (4850, 1782)
        (4900, 1697)
        (4950, 1647)
        (5000, 1714)
  };
  \end{axis}
\end{tikzpicture}
\caption{Distribution of passages lengths in the MS MARCO document corpus.}
\label{figure:ms-marco-doc:corpus_length_distribution}
\end{center}
\end{figure*}
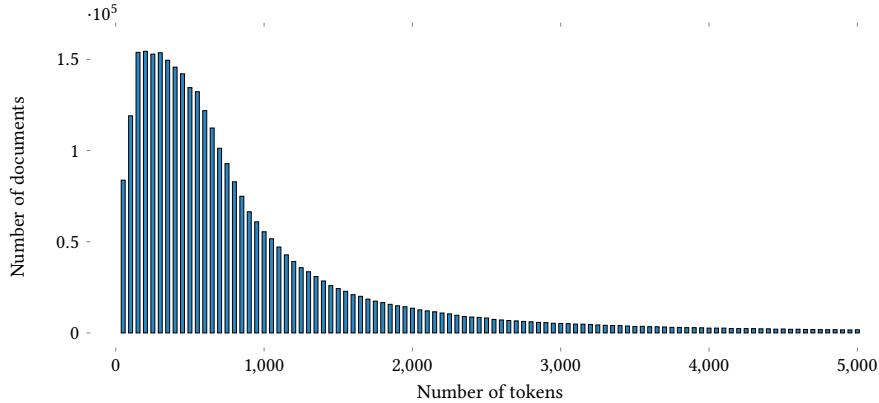

We also evaluated our models on the MS MARCO document ranking dataset, which uses a corpus comprised of 3.2M documents, derived from the MS MARCO passage corpus.
The mean length of each passage is 1131 tokens (median: 584, max: 333757), and the distribution of document lengths\footnote{Also tokenized using Python's string \texttt{split} method.} is shown in Figure~\ref{figure:ms-marco-doc:corpus_length_distribution}.
The general setup of this task is similar to the passage ranking task, with the primary difference being the length of texts in the corpus (passages vs.\ full-length web documents).
We see that the median length of documents is longer than the typical 512 token limit of transformer models, which presents a technical challenge for researchers to overcome.
In total, the dataset contains 367K training queries and 5193 development queries; each query has exactly one relevance judgment.
There are 5793 test queries, but as with the passage ranking task, relevance judgments are only publicly available for the development set queries; obtaining scores to the held-out test set requires a making a submission to the leaderboard.

Similar to the passage ranking task, relevance judgments for the document ranking task are sparse.
Thus, we also evaluated our models on the document ranking task of the TREC 2020 Deep Learning Track~\cite{TREC2020_DL_overview}, which uses the same corpus but has 45 queries with judgements (on a four-point scale), gathered via traditional pooling techniques (with $\sim$200 judgments per query). 

For both the MS MARCO document ranking task and the TREC 2020 Deep Learning Track, we used models trained only on the MS MARCO passage ranking dataset in a zero-shot setting.
That is, we did not use any data from the MS MARCO document ranking dataset to train our models; in addition, no training data from the TREC 2019 Deep Learning Track were used (both passage and document).
However, we are clearly taking advantage of the close relationship between the document and passage datasets, and so it cannot be claimed that we are performing much meaningful transfer learning in this setup.
Our motivation for reusing models from the passage ranking task was to eliminate the need to perform costly fine-tuning and to simplify experimental variables.

\medskip \noindent {\bf ``Expando'' and Keyword Retrieval Settings.}
We performed document expansion on the MS MARCO document corpus with queries predicted by the doc2query-T5 model trained on the MS MARCO passage ranking task.
However, the expansion procedure differed slightly for the document corpus because many documents exceed T5's input length restriction of 512 tokens (unlike with the passage corpus, where all passages fall under the length limit).
To address this issue, we first segmented each document into passages by applying a sliding window of ten sentences with a stride of five.
Each passage was then prepended with the title of the document.
Finally, we performed inference on these passages with doc2query-T5 using a top-$k$ sampling decoder and generated ten queries (i.e., expansions) per passage.

From this, we prepared two separate indexes with Anserini:

\begin{itemize}

\item Per-document expansion.
Each document from the original corpus was appended with all its predicted queries, and each of these augmented documents served as a unit of indexing.

\item Per-passage expansion.
Each passage above, with its predicted queries, served as a unit of indexing.
In this case, each document from the original corpus yielded multiple ``retrievable units'' in the index.
To distinguish among these, we numbered the passages in each document sequentially, and thus each document from the original corpus yielded \texttt{docid\#0}, \texttt{docid\#1}, \texttt{docid\#2}, etc.
This convention made it easy to tell at query time which retrieved passages were from the same underlying document.

\end{itemize}

\noindent For ablation purposes, we also prepared non-expanded counterparts of the above indexes:\ a simple document index and a passage index, segmented in the same manner as above, but without expansions.

In all cases, retrieval was performed with BM25 using Anserini's default parameters $k_1=0.9$, $b=0.4$.
For the TREC 2020 Deep Learning Track topics, we also applied the BM25 + RM3 model (same as in the passage ranking case, using all Anserini default settings).
The results of the keyword queries provided the candidates that fed our mono/duo rerankers, described in detail below.
When evaluating document ranking in isolation, for the per-passage indexes, we constructed a document ranking by retaining the passage from each document with the highest score.
Thus, to obtain a top $k$ document ranking, we retrieve top $k'$ passages from the passage index (where $k' >> k$), and then apply per-document max passage filtering to retain the top $k$ documents.
This simple yet effective approach draws from passage retrieval techniques that date back to the 1990s~\cite{Hearst_SIGIR1993,Callan_SIGIR1994}.

\medskip \noindent {\bf ``Mono'' Settings.}
In the first reranker stage, we simply used the monoT5 model trained on the MS MARCO passage ranking dataset.
However, it is clear from Figure~\ref{figure:ms-marco-doc:corpus_length_distribution} that MS MARCO documents are too long to be directly fed into our models.
Instead, we applied the reranker in two different ways:

For the MS MARCO document ranking task, we used only the per-passage indexes (with and without expansion, for ablation purpose).
At reranking time, we performed monoT5 inference on the top 10000 passages, since each ``retrievable unit'' was, in essence, already pre-segmented and could be directly fed to monoT5.
In other words, we only applied inference to the passages identified by keyword search as being potentially relevant.
As with the passage condition, input to our reranking pipeline did not include the predicted queries.

For the TREC 2020 Deep Learning Track, we only used results from the per-document indexes (with and without expansion, for ablation purposes). 
At reranking time, we first segmented each document into passages using the same technique as above (ten sentence sliding window, five sentence stride, prepending title).
Note that monoT5 was fed passages from the original documents (i.e., without predicted queries). 
We obtained a probability of relevance for each passage by performing inference on it independently, and then selected the highest probability among the passages as the relevance score of the document.
This is similar to the MaxP approach of~\citet{Dai_Callan_SIGIR2019}, although our definitions of passages differ.
In this configuration, the reranker considered the top 1000 keyword search results ($k_0 = 1000$).

\medskip \noindent {\bf ``Duo'' Settings.}
As with monoT5, we used the duoT5 model trained on the MS MARCO passage ranking dataset with default parameters ($k_1 = 50$ and \textsc{SYM-SUM} as the aggregation method).
At reranking time, we used the highest scoring monoT5 passage as the representative passage for each document.
The duoT5 model was fed pairs of representative passages from the documents under consideration to compute the pairwise scores, which were then aggregated to yield the relevance scores of each document.

With duoT5, we increased the maximum input tokens from 512 to 1024 to account for pairs of passages that were longer than the default limit of 512 tokens.
We were able to do so in T5 since the models were trained with relative positional encodings~\cite{shaw-etal-2018-self} and thus can (hopefully) generalize to contexts larger than those seen during training.
This modification, however, imposed additional computational costs that come from the model needing to attend to twice the number of tokens; transformers exhibit quadratic complexity in both time and space with respect to input length~\cite{Kitaev_etal_ICLR2020}.\footnote{Note that increasing the length to 1024 tokens was sufficient in this case. However, for monoT5, such an increase would still not have been sufficient to perform inference on a complete document.}

\subsection{TREC-COVID}
\label{settings:trec-covid}

As one response from the information retrieval community to the global COVID-19 pandemic, the U.S.\ National Institute for Standards and Technology (NIST) organized the TREC-COVID challenge~\cite{TREC-COVID1,TREC-COVID2}, with the goal of evaluating systems that help stakeholders access reliable scientific evidence about the virus.
The series of evaluations, which ran from mid-April to late-August 2020, used the COVID-19 Open Research Dataset (CORD-19)~\cite{Wang:2004.10706:2020}, curated by the Allen Institute for AI (AI2).

Due to the evolving nature of both the scientific literature and information needs, the evaluation was organized into a series of ``rounds'' (five in total), each of which used CORD-19 at a snapshot in time.
Each round essentially comprised a separate TREC-style evaluation, with both new information needs (i.e., topics) as well as holdover information needs from previous rounds; the first round had 35 topics, and each subsequent round added five additional, yielding a set of 50 topics by round 5.
One example is ``serological tests that detect antibodies of COVID-19''.
For each round, relevance judgments were gathered using standard pooling methodology from participant runs.
Since each round used a corpus that was, to an approximation, a superset of the previous (reflecting the growing literature),\footnote{While this characterization is conceptually accurate, articles were both added and removed in each round. There were also significant complexities involving near-duplicate documents in the corpus and unstable document identifiers, for example, reflecting cases where a preprint was later published in a peer-reviewed venue.} NIST adopted a residual evaluation methodology.
That is, at each round, all previously judged documents were removed entirely from consideration (this was operationalized by removing the docids of previously judged documents from all new run submissions).

Our participation in TREC-COVID is detailed in~\citet{ZhangEdwin_etal_SDP2020}; here we focus only on the aspects that are relevant to our Expando-Mono-Duo design pattern, as opposed to a general overview of our team's efforts.

\medskip \noindent {\bf ``Expando'' and Anserini Settings.}
For TREC-COVID, we used the model trained for the MS MARCO passage ranking task.
That is, document expansion was performed in a zero-shot manner, in that the model had not been fine-tuned on data from the target corpus (CORD-19).

Due to limited computational resources, we only generated expansions from the article abstracts.
However, even the abstracts alone often exceeded T5's input length restriction of 512 tokens.
To address this issue, we used the same sliding window approach described in Section~\ref{sec:marco_doc_setup} (window of ten sentences with a stride of five, passages prepended with the title of the article).
Inference was performed on these passages using a top-$k$ sampling decoder that generated 40 queries (i.e., expansions) per abstract passage.
Interestingly, based on manual examination of the predicted queries, we found their quality to be quite good, at least to untrained experts.
This was a surprising finding given the specialized domain:\ while it is true that T5 had been exposed to scientific text during pretraining, the model was {\it not} fine-tuned with any data that could be considered ``in domain''.

Previous work on searching full-text articles~\cite{Lin_BMCBioinformatics2009} and results from the first round of TREC-COVID showed value in fusion approaches that combined multiple sources of evidence.
Thus, starting in round 2, we adopted an approach that fused results from three different indexes:

\begin{itemize}

\item An index where each ``document'' is comprised of the title and abstract of an article in the corpus.

\item An index where each ``document'' is comprised of the full text of an article in the corpus (including title and abstract).
For articles in the corpus that do not have full text, the ``document'' contains only the title and abstract.

\item A paragraph-level index structured as follows:\ each full-text article was segmented into paragraphs, based on the markup provided in the corpus, and for {\it each} paragraph, we created a ``document'' comprising the title, abstract, and that paragraph.
The title and abstract alone comprised an additional ``document''.
Thus, a full-text article with $n$ paragraphs yielded $n+1$ separate retrieval units.
With this index, a query is likely to retrieve multiple paragraphs from the same article; we selected the highest-scoring paragraph as the score of that article.

\end{itemize}

\noindent First-stage retrieval (i.e., the input to our reranking pipeline) comprised results combined from all three indexes using reciprocal rank fusion~\cite{cormack2009reciprocal}.

There is an additional complexity that needs explaining:\ one of the lessons learned in the first round of TREC-COVID was the importance of query formulation.
TREC-COVID information needs (i.e., topics) were modeled after standard {\it ad hoc} retrieval topics, comprising ``query'', ``question'', and ``narrative'' fields.\footnote{These fields parallel the ``title'', ``description'', and ``narrative'' structure of standard TREC topics.}
Each of the fields described the information needs in increasing detail:\ a few keywords, a sentence, and a short paragraph, respectively.
Of the three fields, taking the combination of the ``query'' and ``question'' (i.e., concatenating the contents of both fields) was found to be the most effective based on empirical results from round~1 (which is consistent with the literature).

Beyond this simple approach, researchers from the University of Delaware discovered an effective technique for generating queries from the TREC-COVID topics, based on results from round 1.
They began with non-stopwords from the query field and then further added named entities extracted from the question field using ScispaCy~\cite{neumann-etal-2019-scispacy}.
Starting in round 2, keyword queries generated using this technique (dubbed the UDel query generator) were shared with all participants {\it prior} to each round's deadline.
Since the query generator was not further modified, it is accurate to claim that this technique was used in rounds 2 through 5 in a blind manner.

For TREC-COVID, there were thus two commonly adopted query generation approaches.
Combined with the three different index conditions discussed above, this led to two different fusion runs:

\begin{itemize}

\item \run{fusion1}: reciprocal rank fusion of results from the abstract, full-text, and paragraph indexes with the ``query'' and ``question'' fields (concatenated together) as the search query.

\item \run{fusion2}: reciprocal rank fusion of results from the abstract, full-text, and paragraph indexes with queries generated by the UDel query generator.

\end{itemize}

\noindent Starting in round 2, these two fusion runs, \run{fusion1} and \run{fusion2}, were provided as community baselines by our team, made available for downloading prior to the deadline of each round so that participants could build on them (and indeed many did).
The experimental runs we describe in this paper took \run{fusion1} and \run{fusion2} as independent first-stage retrievers, reranked their output, and then combined their results using reciprocal rank fusion (details in Section~\ref{results:trec-covid}).

Document expansion was applied in the following manner:\ for each of the index conditions above, we expanded the ``documents'' by appending all the expansion queries to form a total of three more augmented indexes enhanced by doc2query-T5.
First-stage retrieval then used these augmented indexes, exactly as before (i.e., reciprocal rank fusion, as described above).
As with the other tasks described in this section, downstream rerankers consumed the original text, i.e., we did {\it not} feed any of the expansion queries into the model for inference.
The same two query generation approaches described above could also be applied, yielding variants of \run{fusion1} and \run{fusion2} based on the augmented indexes.

In summary, there are four possible inputs to our reranking pipeline:\ results from \run{fusion1}, \run{fusion2} and the corresponding expanded versions based on doc2query-T5.
We will further clarify the exact experimental conditions evaluated in Section~\ref{results:trec-covid} when we present results.

\medskip \noindent {\bf ``Mono'' Settings.}
We used the monoT5 model fine-tuned on the MS MARCO passage dataset and then fine-tuned again on Med-MARCO, which is a subset of the MS MARCO passage dataset where only queries containing medical terms are kept~\cite{macavaney2020sledge}.
For this second fine-tuning, we trained for 1K steps using batches of 128 examples and a learning rate of $10^{-4}$.
Although Med-MARCO is ten times smaller than the full MS MARCO passage dataset in terms of (query, relevant passage) pairs for training, we noticed a small gain in effectiveness when fine-tuning on it (additionally).

To be clear, the only adaptation of monoT5 for TREC-COVID was the additional fine-tuning with Med-MARCO.
Specifically, the model was not provided (for fine-tuning or otherwise) task-specific data (e.g., relevance judgments), even though such data became available in the later rounds.

At reranking time, we only considered article titles and abstracts from CORD-19, primarily due to length limitations already discussed.
We used the same sliding window approach described above, and used the maximum score of a passage as the overall score of that article, just as with our MS MARCO document ranking experiments.
Only the ``question'' field of the topic was used for the query text $q$ in the input sequence template of our rerankers.

\medskip \noindent {\bf ``Duo'' Settings.}
Similar to monoT5, we used duoT5 trained on the MS MARCO passage dataset and then fine-tuned again on Med-MARCO.
In this second fine-tuning, we trained for 1K steps using batches of 128 examples and a learning rate of $10^{-4}$.
We used the default parameters ($k_1=50$ and $\textsc{Sym-Sum}$ as the pairwise aggregation method).
At reranking time, we used the highest scoring monoT5 passage (eight sentence sliding window, four sentence stride, prepending title) as the representative passage for each document.
Like the MS MARCO document ranking task, we increased the maximum input tokens to 1024 to account for pairs of passages that were longer than the default limit of 512 tokens.

\section{Results}

We present experimental results from applying our Expando-Mono-Duo pattern to five {\it ad hoc} retrieval tasks in different domains.
Empirically, our approach has been validated to be at or near the state of the art across all of these tasks.

\subsection{MS MARCO Passage Ranking}
\label{sec:marco_passage_results}

For the MS MARCO passage ranking task, Table~\ref{tab:duot5} reports MRR@10, the official metric, for various combinations of our techniques, as shown in the $H_{-1}$, $H_{0}$, $H_{1}$, $H_{2}$ columns:\ these provide ablations to show individual component contributions.
Note that obtaining a score on the test set requires submitting a run to the leaderboard organizers, which we did not do for less interesting conditions.
Overly frequent submissions to the leaderboard are actively discouraged by the organizers, and we wished to minimize ``probes'' of the blind held out test set.
Thus, for some of the ablation and contrastive conditions, we only report scores on the development set.

\begin{table*}[t]
\begin{center}
\begin{small}
\begin{tabular}{lllll|cc}
\toprule
\multicolumn{5}{l|}{Model} & Dev & Test\\
\toprule
& \multicolumn{4}{c|}{Expand-Mono-Duo Variants} & \\
& $H_{-1}$ & $H_{0}$ & $H_{1}$ & $H_{2}$ \\
\midrule
(1) & - & BM25 & - & - & 0.184 & 0.184 \\
(2) & - & BM25 & monoT5-base & - & 0.381 & - \\
(3) & - & BM25 & monoT5-large & - & 0.393 & - \\
(4) & - & BM25 & monoT5-3B & - & 0.398 & - \\
\midrule
(5) & doc2query-T5 & BM25 & - & - & 0.277 & 0.272\\
(6) & doc2query-T5 & BM25 & monoT5-3B & - & 0.409 & - \\
(7) & doc2query-T5 & BM25 & monoT5-3B & duoT5-3B & 0.420 & 0.408 \\
\midrule
(8) & - & BM25 & monoT5-base & duoT5-base & 0.392 & - \\
(9) & - & BM25 & monoT5-3B & duoT5-base & 0.409 & - \\
\bottomrule
(10) &\multicolumn{4}{l|}{BM25 + BERT-large  \cite{Nogueira_etal_arXiv2019_multistageBERT}} & 0.372 & 0.365 \\
(11) &\multicolumn{4}{l|}{TFR-BERT Ensemble \cite{han2020learningtorank}} & 0.405 & 0.395 \\
(12) & \multicolumn{4}{l|}{RocketQA \cite{Qu:2010.08191:2020}} & 0.439 & 0.426 \\ 
\bottomrule
\end{tabular}
\end{small}
\end{center}
\caption{Results on the MS MARCO passage ranking task, showing the official metric MRR@10 on the development and test sets. Note that scores on the test set are only available via a leaderboard submission.}
\label{tab:duot5}
\end{table*}

At the time of submission (May 2020), the full Expando-Mono-Duo configuration, condition (7), was the best result atop the very competitive leaderboard (which had received around one hundred submissions at the time).
For comparison, we provide scores from two other runs:\
BM25 + BERT-large, condition (10), can be characterized as the ``baseline'' of transformer-based techniques~\cite{Lin_etal_arXiv2020_ptr4tr}.
The reported effectiveness is copied from~\citet{Nogueira_etal_arXiv2019_multistageBERT}, but the general approach dates back to~\citet{Nogueira:1901.04085:2019}.
TFR-BERT~\cite{han2020learningtorank}, condition (11), was the best model just prior to our leaderboard submission.
As of January 2021, our submission currently ranks fourth on the leaderboard; the top spot is occupied by RocketQA~\cite{Qu:2010.08191:2020}, condition (12).
While our run is no longer the best known result, Expando-Mono-Duo remains near the state of the art in terms of effectiveness on this task.
Since our submission to the MS MARCO passage leaderboard in May 2020, we have not worked on this task further---to, for example, incorporate more recent innovations that might bolster the effectiveness of Expando-Mono-Duo.
It is possible that recent advances are orthogonal (and hence cumulatively beneficial) to the techniques that we propose here.

Conditions (2), (3), and (4) focus on single stage reranking using monoT5, without document expansion; these differ in the size of the underlying T5 model (monoT5-base vs.\ monoT5-large vs.\ monoT5-3B).
It is apparent that increasing model sizes yields improvements in effectiveness, as expected.
These results show that T5 is more effective than BERT, even factoring in model size differences, bolstering the arguments of~\citet{Nogueira_etal_FindingsEMNLP2020} in advocating ranking with sequence-to-sequence models.
In fact, a single large model, monoT5-3B, condition (4), approaches TFR-BERT, which is an ensemble of 15 BERT-large models that reranked the top 1000 passages from both BM25 and DeepCT~\cite{dai2019contextaware}.

Conditions (5), (6), and (7) include document expansion, using models based on T5-3B  for reranking.
We see that adding document expansion (stage $H_{-1}$) to enhance first-stage retrieval (stage $H_0$), condition (5) vs.~(1), yields a large boost on the test set (a nearly 50\% relative improvement); these gains carry over to the held-out test set.
Note that this gain comes without the need to perform any (computationally expensive) neural inference at query time, although first-stage retrieval latency increases modestly since the expanded documents are longer.

On top of doc2query-T5, adding monoT5-3B, condition (6), and duoT5-3B, condition (7), each contribute additional cumulative gains, and the combination gives us the best score, corresponding to our best entry on the leaderboard, condition (7).
Each of these gains are statistically significant, based on the $t$-test (at $p<0.01$), on the development set (the only split we have access to).
In other words, every component in our multi-stage ranking pipeline contributes significantly to end-to-end effectiveness.

Condition (8) represents an ablation study that is intended to be contrasted with condition (4), designed to answer the question:\ What's more important, the size of the sequence-to-sequence model, or the monoT5/duoT5 reranking design?
The results are clear:\ a single stage reranker with a larger model beats the pointwise/pairwise combination with a smaller model.
However, as the comparison with condition (2) clearly shows, the addition of ``Duo'' improves over ``Mono'' alone, even with the smaller T5-base model.
Finally, condition (9) shows that we need a ``Duo'' that is at least as large as ``Mono'' to be effective.
That is, if ``Duo'' uses a smaller model than ``Mono'', it does not appear to improve over ``Mono'' alone, see condition (6) vs.\ condition (9).

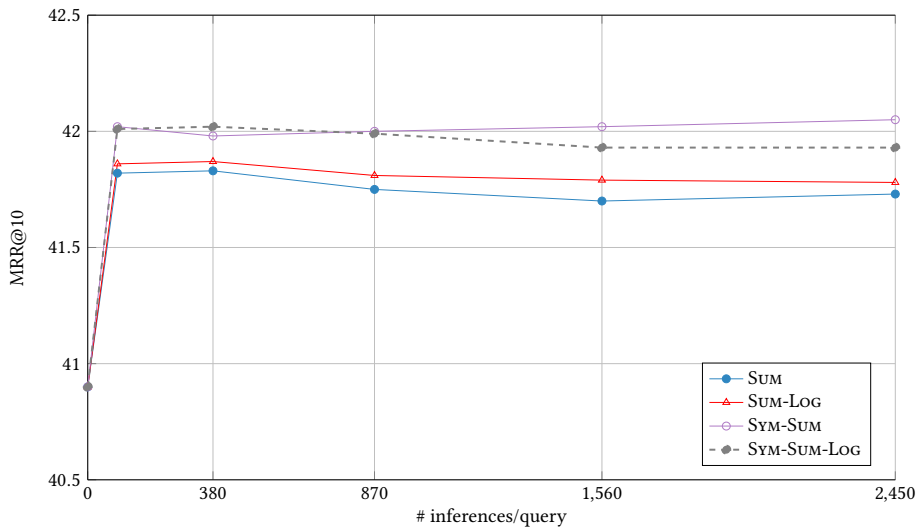
\begin{figure}[t]
\centering
\begin{tikzpicture}[scale = 0.75]
\begin{axis}[
width=1.00\columnwidth,
height=0.618\columnwidth, 
legend cell align=left,
mark options={mark size=3},
font=\normalsize,
axis y line*=left,
xmin=0, xmax=2450,
ymin=40.5, ymax=42.5,
log ticks with fixed point,
xtick={0, 380, 870, 1560, 2450},
ytick={40.5, 41.0, 41.5, 42.0, 42.5},
legend pos=south east,
xmajorgrids=true,
ymajorgrids=true,
xlabel style={font = \normalsize, yshift=1ex},
xlabel=\# inferences/query,
ylabel= MRR@10,
ylabel style={font = \normalsize, yshift=0ex}]
    \addplot[mark=*,g-blue, mark options={scale=1}] plot coordinates {
    (0, 40.9)(90, 41.82)(380, 41.83)(870, 41.75)(1560, 41.70)(2450, 41.73)
    };
    \addlegendentry{\textsc{Sum}}
    \addplot[mark=triangle,red, mark options={scale=1}] plot coordinates {
    (0, 40.9)(90, 41.86)(380, 41.87)(870, 41.81)(1560, 41.79)(2450, 41.78)
    };
    \addlegendentry{\textsc{Sum-Log}}
    \addplot[mark=o,g-purple, mark options={scale=1}] plot coordinates {
     (0, 40.9)(90, 42.02)(380, 41.98)(870, 42.00)(1560, 42.02)(2450, 42.05)
    };
    \addlegendentry{\textsc{Sym-Sum}}
    \addplot[line width=1pt, dashed, mark=*,gray, mark options={scale=1}] plot coordinates {
    (0, 40.9)(90, 42.01)(380, 42.02)(870, 41.99)(1560, 41.93)(2450, 41.93)
    };
    \addlegendentry{\textsc{Sym-Sum-Log}}
\end{axis}
\end{tikzpicture}
\caption{Number of inferences per query vs.\ the effectiveness of duoT5-3B on the MS MARCO passage development set when varying the number of candidates $k_1$. We evaluated at $k_1=\{0, 10, 20, 30, 40, 50\}$, where $k_1=0$ corresponds to using only monoT5-3B. The values in the {\it x}-axis are computed as $k_1 \times (k_1 - 1)$.}
\label{fig:duot5_cutoff}
\end{figure}

In Figure~\ref{fig:duot5_cutoff}, we examine the effectiveness of duoT5-3B on the development set when reranking different numbers of candidates $k_1$ from the output of monoT5-3B.
In these experiments, we considered $k_1=\{0, 10, 20, 30, 40, 50\}$, where $k_1=0$ corresponds to using only monoT5-3B; in all cases, $k_0$ is set to 1000, i.e., monoT5 considers the top 1000 candidates from first-stage retrieval.
The values in the {\it x}-axis are computed as $k_1 \times (k_1 - 1)$, which is the total number of inferences needed at the duo stage.
We find that \textsc{Sym-Sum} and \textsc{Sym-Sum-Log} performs best, indicating that combining scores from both $(d_i, d_j)$ and $(d_j, d_i)$ improves reranking.
Interestingly, most of the impact of pairwise reranking comes from a small value of $k_1$, which means that the gains in MRR come from reshuffling results in the top ranks.
In fact, we see that in all but \textsc{Sym-Sum}, model effectiveness drops slightly as we increase $k_1$.
Note that our leaderboard submissions used $k_1=50$ with \textsc{Sym-Sum}.

\subsection{TREC 2020 Deep Learning Track Passage Ranking}
\label{sec:trecdl_passage_results}

Table~\ref{tab:results_dl_pr} presents results from the passage ranking condition of the TREC 2020 Deep Learning Track; the columns $H_{-1}$, $H_{0}$, $H_{1}$, $H_{2}$ denote different settings of our Expand-Mono-Duo design.
All of these results represent official submissions to the evaluation.
It is worthwhile to emphasize that the underlying models did {\it not} take advantage of relevance judgments from the TREC 2019 Deep Learning Track.
Our best run in terms of nDCG@10, condition (5), was the second best run (on a per-team basis) submitted to the evaluation.
The first block of the table, conditions (1)--(3), presents results without document expansion, and the second block of the table, conditions (4)--(7), presents results with doc2query-T5.

Let us first examine the impact of query expansion using pseudo-relevance feedback.
The first two rows, conditions (1)~and~(2), represent standard bag-of-words baselines, with BM25 and BM25 + RM3, respectively (see Section~\ref{sec:setup:passage}).
As expected, pseudo-relevance feedback (BM25 + RM3) increases effectiveness, which is consistent with decades of information retrieval research.
Note that confirmation of this finding is possible only with the ``dense'' judgments available in TREC.
As demonstrated by~\citet{Lin_etal_arXiv2020_ptr4tr}, with the ``sparse'' judgments from the MS MARCO passage dataset, BM25 + RM3 actually scores lower, due to the inability of the MS MARCO passage judgments (on average, only one relevant passage per query) to properly capture (i.e., reward) the effects of query expansion.
We similarly observe the beneficial effects of RM3 when applied on top of the doc2query-T5 expanded index, condition (4) vs.\ condition (6), in terms of MAP, MRR, and R@1K, but not in terms of nDCG@10.

\begin{table*}[t]
\begin{small}
\begin{center}
\begin{tabular}{lllll|llHll}
\toprule
& \multicolumn{4}{c|}{Expand-Mono-Duo Variants} & \\
& $H_{-1}$ & $H_{0}$ & $H_{1}$ & $H_{2}$ & MAP & nDCG@10 & nDCG@1K & MRR & R@1K\\
\toprule
(1) & - & BM25 & - & - & 0.2856 & 0.4796 & 0.5830 & 0.6585 & 0.7863 \\
(2) & - & BM25 + RM3 & - & - & 0.3019 & 0.4821 & 0.6046 & 0.6360 & 0.8217\\
(3) & - & BM25 + RM3 & monoT5-3B & duoT5-3B & 0.5355 & 0.7583 & 0.7387 & 0.8759 & 0.8217 \\
\midrule
(4) & doc2query-T5 & BM25 & - & - & 0.4074 & 0.6187 & 0.6840 & 0.7326 & 0.8452 \\
(5) & doc2query-T5 & BM25 & monoT5-3B & duoT5-3B & 0.5609 & 0.7837 & 0.7539 & 0.8798 & 0.8452 \\
(6) & doc2query-T5 & BM25 + RM3 & - & - & 0.4295 & 0.6172 & 0.7041 & 0.7424 & 0.8699 \\
(7) & doc2query-T5 & BM25 + RM3 & monoT5-3B & duoT5-3B  & 0.5643 & 0.7821 & 0.7732 & 0.8798 & 0.8699 \\
\bottomrule
\end{tabular}
\end{center}
\end{small}
\caption{Results on the TREC 2020 Deep Learning Track Passage Ranking Task.}
\label{tab:results_dl_pr}
\end{table*}

Conditions (2) and (3), (4) and (5), and (6) and (7) form contrastive minimal pairs with and without mono/duoT5 reranking using T5-3B.
Note that these pairs have the same recall, since for each pair, the latter reranks output from the former and thus cannot find any additional relevant documents.
In each case, we see that our mono/duoT5 reranking pipeline yields a large increase in effectiveness in all first-stage configurations:\ BM25 + RM3, BM25 with doc2query-T5 expansion, and BM25 + RM3 with doc2query-T5 expansion.
These findings are consistent with the results on the MS MARCO passage dataset.
Conditions (5) and (7) demonstrate that while pseudo-relevance feedback improves  recall, it is not clear if it improves end-to-end effectiveness in terms of metrics like MAP, nDCG@10, and MRR after reranking.
In other words, in a pipeline with doc2query-T5, monoT5, and duoT5, it is unclear if pseudo-relevance feedback is still necessary.

Here, we see an interesting relationship between document expansion and pseudo-relevance feedback (which is query expansion) from the end-to-end perspective (with reranking).
Put differently, the comparison between conditions (5) and (7) suggests that with document expansion, query expansion does not seem to matter much.
In fact, our highest nDCG@10 score, condition (5), and also the best submission to the TREC 2020 Deep Learning Track, did {\it not} use query expansion.
However, the comparison between conditions (3) and (7) suggests that with query expansion, document expansion still helps, i.e., the latter beats the former in all metrics.
Furthermore, between the two techniques separately, document expansion alone, condition (5), is more effective than query expansion alone, condition (3).
Since BM25 + RM3 is only a single query expansion technique (and does not exploit neural networks), this is not an entirely fair comparison between document expansion and query expansion methods {\it in general}.
However, we do note that document expansion techniques can take advantage of longer texts as input (hence more context) than query expansion techniques, since queries are usually much shorter.
These interesting observations deserve more study in future work.

\subsection{MS MARCO Document Ranking}
\label{sec:marco_document_results}

For the MS MARCO document ranking task, Table~\ref{tab:doc_duot5} reports MRR@100, the official metric, for certain combinations of our techniques, as shown in the $H_{-1}$, $H_{0}$, $H_{1}$, $H_{2}$ columns:\ these provide ablations to show individual component contributions.
As with the passage ranking task, obtaining a score on the test set requires submitting a run to the leaderboard, and the evaluation organizers actively discourage submission of runs that are ``too similar''.
Hence, we focused on more interesting experimental conditions and refer readers to the more thorough ablation analysis on the MS MARCO passage ranking dataset in Section~\ref{sec:marco_passage_results}.

\begin{table*}[t]
\begin{center}
\begin{small}
\begin{tabular}{lllll|ll}
\toprule
\multicolumn{5}{l|}{Model} & Dev & Test\\
\toprule
& \multicolumn{4}{c|}{Expand-Mono-Duo Variants} & \\
& $H_{-1}$ & $H_{0}$ & $H_{1}$ & $H_{2}$ \\
\toprule
(1) & - & BM25 & - & - & 0.268 & - \\
(2) & doc2query-T5 & BM25 & - & - & 0.318 & 0.284 \\
(3) & doc2query-T5 & BM25 & monoT5-3B & - & 0.411 & 0.362 \\
(4) & doc2query-T5 & BM25 & monoT5-3B & duoT5-3B & 0.426 & 0.370 \\
\midrule
(5) &\multicolumn{4}{l|}{PROP\_step400K base (ensemble v0.1)} & 0.455 & 0.401 \\
\bottomrule
\end{tabular}
\end{small}
\end{center}
\caption{Results on the MS MARCO document ranking task, showing the official metric MRR@100 on the development and test sets. Note that scores on the test set are only available via a leaderboard submission.}
\label{tab:doc_duot5}
\end{table*}

Note that in condition (1), the only one that doesn't use doc2query-T5, we used the per-passage index while in the rest of the conditions, we used the per-passage expansion index, both as described in Section~\ref{sec:marco_doc_setup}.
For the reranking conditions, conditions (3) and (4), the top 10000 passages from keyword search were processed (compared to top 1000 for the passage ranking task).
Thus, these runs required a significantly larger (around 10 times larger) compute budget compared to the MS MARCO passage ranking runs.

At the time of submission (September 2020), the full Expando-Mono-Duo configuration, condition (4), was the best result atop the very competitive leaderboard.
The best run on the leaderboard (as of January 2021) is shown as condition (5) in Table~\ref{tab:doc_duot5}.
While it is no longer the best known result, Expando-Mono-Duo remains near the state of the art in terms of effectiveness on this task especially given that our pipeline is zero-shot.
In contrast, most of the top submissions involve ensembles and training directly on the MS MARCO document ranking dataset.
As with the passage ranking leaderboard, we have not worked on this task since our submission, and it is possible that recent innovations can be incorporated into our approach to further improve the effectiveness of Expand-Mono-Duo.

We see that adding document expansion (stage $H_{-1}$) to enhance first-stage retrieval (stage $H_0$), condition (2) vs.~(1), yields a boost on the development set.
This is consistent with results from the passage ranking task.
Note here that, as with the passage ranking task, this gain comes at only a modest increase in first-stage retrieval latency since the expanded documents are longer; computationally expensive neural inference is not required at query time.

On top of doc2query-T5, adding monoT5-3B, condition (3), and duoT5-3B, condition (4), each contributes additional cumulative gains, and the combination gives us the best score, corresponding to our best entry on the leaderboard, condition (4).
Each of these gains are statistically significant, based on the $t$-test (at $p<0.01$), on the development set (note that we do not have access to the test set).
In other words, every component in our multi-stage ranking pipeline contributes significantly to end-to-end effectiveness.
Once again, these findings are consistent with results on the MS MARCO passage ranking task.

\subsection{TREC 2020 Deep Learning Track Document Ranking}
\label{sec:trecdl_document_results}

Table~\ref{tab:results_dl_dr} presents results from the document ranking condition of the TREC 2020 Deep Learning Track; the columns $H_{-1}$, $H_{0}$, $H_{1}$, $H_{2}$ denote different settings of our Expand-Mono-Duo setting.
All of these results represent official submissions for evaluation.
It is worthwhile to emphasize that the underlying models did {\it not} take advantage of relevance judgments from the TREC 2019 Deep Learning Track.
Our best run in terms of nDCG@10, condition (5), was the best run submitted to TREC 2020.
The first block of the table, conditions (1)--(3), presents results without document expansion.
Due to limits on the number of submissions allowed, our runs for conditions (4)--(7) only used per-document expansions.

\begin{table*}[t]
\begin{center}
\begin{small}
\begin{tabular}{lllll|llHll}
\toprule
& \multicolumn{4}{c|}{Expand-Mono-Duo Variants} & \\
& $H_{-1}$ & $H_{0}$ & $H_{1}$ & $H_{2}$ & MAP & nDCG@10 & NDCG@1K & MRR & R@1K \\
\midrule
(1) & - & BM25 & - & - & 0.3791 & 0.5271 & 0.5647 & 0.8521 & 0.8085 \\
(2) & - & BM25 + RM3 & - & - & 0.4006 & 0.5248 & 0.5726 & 0.8541 & 0.8260 \\
(3) & - & BM25 + RM3 & monoT5-3B & duoT5-3B & 0.5270 & 0.6794 & 0.6929 & 0.9476 & 0.8260 \\
\midrule
(4) & doc2query-T5 & BM25 & - & - & 0.4230 & 0.5885 & 0.6115 & 0.9369 & 0.8403 \\
(5) & doc2query-T5 & BM25 & monoT5-3B & duoT5-3B & 0.5422 & 0.6934 & 0.7089 & 0.9476 & 0.8403\\
(6) & doc2query-T5 & BM25 + RM3 & - & - & 0.4228 & 0.5407 & 0.5902 & 0.8147 & 0.8596 \\
(7) & doc2query-T5 & BM25 + RM3 & monoT5-3B & duoT5-3B & 0.5427 & 0.6900 & 0.7122 & 0.9476 & 0.8596 \\
\bottomrule
\end{tabular}
\end{small}
\end{center}
\caption{Results on the TREC 2020 Deep Learning Track Document Ranking Task.}
\label{tab:results_dl_dr}
\end{table*}

The first two rows represent standard bag-of-words baselines, with BM25 and BM25 + RM3, respectively.
This parallels the conditions for the passage ranking task.
Pseudo-relevance feedback increases effectiveness in terms of R@1K, the most important metric for first-stage retrieval since it sets the effectiveness upper bound for the entire reranking pipeline.
We see that MAP increase as well, but nDCG@10 and MRR are essentially unchanged (since they are not recall-oriented metrics).
The finding here is consistent with results from the passage ranking task.
The comparison between condition (4) and condition (6) illustrates the effects of applying pseudo-relevance feedback on top of doc2query-T5 (without reranking).
We see that R@1K improves, but the other metrics are either essentially unchanged or actually degrade.
Thus, the gains from doc2query-T5 and RM3 do {\it not} appear to be additive.

Note that conditions (2) and (3), (4) and (5), and (6) and (7) have the same recall, since for each pair, the latter reranks output from the former and thus cannot find any additional relevant documents.
However, for each of the pairs, the mono/duo reranking pipeline substantially increases effectiveness.
These findings are consistent with previous results.

As with the passage ranking case, the comparisons between conditions (3), (5), and (7) illustrate the effects of document expansion vs.\ query expansion (pseudo-relevance feedback with RM3) in an end-to-end setting (with full mono/duoT5 reranking).
Comparing conditions (5) and (7), we see that with document expansion, query expansion doesn't add much.
However, starting with query expansion, document expansion improves effectiveness; this is seen by comparing condition (3) and condition (7).
Individually, document expansion (alone) appears to be more effective than query expansion (alone), based on comparing condition (3) and condition (5).
These findings are consistent with results in Section~\ref{sec:trecdl_passage_results} on passage ranking.

\subsection{TREC-COVID}
\label{results:trec-covid}

\begin{table*}[t]
\begin{center}
\begin{small}
\begin{tabular}{lHllHrrrH}
\toprule
& Team & Run & Description & Type & nDCG@10 & P@5 & MAP & Judged@5 \\
\toprule
\multicolumn{3}{l}{\textbf{Round 1}: 30 topics} \\
(1a) &sabir & \run{sabir.meta.docs} &  &automatic & 0.6080 & 0.7800 & 0.3128 & 1.0000\\
(1b) &covidex & \run{T5R1} & monoT5-3B & automatic & 0.5223 & 0.6467 & 0.1919 & 1.0000\\
\midrule
\multicolumn{3}{l}{\textbf{Round 2}: 35 topics} \\
(2a) &GUIR\_S2 & \run{GUIR\_S2\_run1} & & automatic & 0.6251 & 0.7486 & 0.2842 & 0.9885 \\
(2b) &covidex & \run{covidex.t5} & monoT5-3B & automatic & 0.6250 & 0.7314 & 0.2880 & 1.0000 \\
(2c) &anserini & \run{r2.fusion2} & & automatic & 0.5553 & 0.6800 & 0.2725 & 1.0000 \\
(2d) &anserini & \run{r2.fusion1} & & automatic & 0.4827 & 0.6114 & 0.2418 & 1.0000 \\
\midrule
\multicolumn{3}{l}{\textbf{Round 3}: 40 topics} \\
(3a) &SFDC & \run{SFDC-fus12-enc23-tf3} &  & automatic & 0.6867 & 0.7800 & 0.3160 & 0.9850 \\
(3b) &covidex & \run{r3.duot5} & monoT5-3B + duoT5-3B & automatic & 0.6626 & 0.7700 & 0.2676 & 0.8850 \\
(3c) &covidex & \run{r3.monot5} & monoT5-3B & automatic & 0.6596 & 0.7800 & 0.2635 & 0.9300 \\
(3d) &anserini & \run{r3.fusion2} & & automatic & 0.6100 & 0.7150 & 0.2641 & 0.9550 \\
(3e) &anserini & \run{r3.fusion1} & & automatic & 0.5359 & 0.6100 & 0.2293 & 0.8950 \\
\toprule
& Team & Run & Description & Type & nDCG@20 & P@20 & AP & Judged@5 \\
\toprule
\multicolumn{3}{l}{\textbf{Round 4}: 45 topics} \\
(4a) &covidex & \run{covidex.r4.d2q.duot5} & doc2query-T5 + monoT5-3B + duoT5-3B & automatic & 0.7219 & 0.7267 & 0.3122 \\
(4b) &covidex & \run{covidex.r4.duot5} & monoT5-3B + duoT5-3B & automatic & 0.6877 & 0.6922 & 0.3283 \\
(4c) &uogTr & \run{uogTrDPH\_QE\_SCB1} & & automatic & 0.6820 & 0.7144 & 0.3457 \\
(4d) &anserini & \run{r4.fusion2} & & automatic & 0.6089 & 0.6589 & 0.3088 \\
(4e) &anserini & \run{r4.fusion1} & & automatic & 0.5244 & 0.5611 & 0.2666 \\
\midrule
\multicolumn{3}{l}{\textbf{Round 5}: 50 topics} \\
(5a) &covidex & \run{covidex.r5.d2q.2s} & doc2query-T5 + monoT5-3B + duoT5-3B & automatic & 0.7539 & 0.7700 & 0.3227 \\
(5b) &covidex & \run{covidex.r5.2s} & monoT5-3B + duoT5-3B & automatic & 0.7457 & 0.7610 & 0.3212 \\
(5c) &uogTr & \run{uogTrDPH\_QE\_SB\_CB} & & automatic & 0.7427 & 0.7910 & 0.3305 \\
(5d) &covidex & \run{covidex.r5.d2q.1s} & doc2query-T5 + monoT5-3B & automatic & 0.7121 & 0.7320 & 0.3150 \\
(5e) &covidex & \run{covidex.r5.1s} & monoT5-3B & automatic & 0.6960 & 0.7070 & 0.3119 \\
(5f) &anserini & \run{r5.fusion2} & & automatic & 0.6007 & 0.6440 & 0.2734 \\
(5g) &anserini & \run{r5.fusion1} & & automatic & 0.5313 & 0.5840 & 0.2314 \\
\bottomrule
\end{tabular}
\end{small}
\end{center}
\vspace{0.25cm}
\caption{Selected TREC-COVID results. Rows show combinations of Expando-Mono-Duo as well as baselines and submissions from other teams for comparison. Note that the metrics used in the first three rounds are different from those used in the final two rounds.}
\label{tab:covid-results}
\end{table*}

Official results from TREC-COVID are shown in Table~\ref{tab:covid-results}.
The official evaluation metric was nDCG, at rank cutoff 10 for the first three rounds, increased to 20 for rounds 4 and 5.
NIST also reported a few other metrics, including precision at a fixed ranked cutoff and average precision (AP) to the standard rank depth of 1000.
It is worth emphasizing that due to the residual collection evaluation methodology, and the fact that the corpora and topics were different, scores across rounds {\it are not} comparable.

A thorough discussion of all results from the TREC-COVID challenge is obviously beyond the scope of this paper, and thus we focus only on runs that directly pertain to the Expando-Mono-Duo pattern.
Table~\ref{tab:covid-results} shows only ``automatic'' runs that were formally submitted to the evaluation.
In an automatic run, manual intervention was not allowed.
This contrasted with two other categories of runs that were accepted: ``feedback'' runs, which could leverage relevance judgments from previous rounds but otherwise could not involve human intervention, and ``manual'' runs, where any human intervention was allowed.
These run categories are interesting as well, but are not germane to the focus of our work.

The ``Run'' column of Table~\ref{tab:covid-results} shows the run identifier for readers interested in matching our submissions with official results, and the ``Description'' column provides a short description in the Expando-Mono-Duo context.
In rounds 1, 2, and 3, we present our own runs as well as the best automatic run from that round---these are noted as conditions (1a), (2a), and (3a), respectively.
In rounds 2 and 3, we submitted the second-best runs (on a per-team basis).
In rounds 4 and 5, we submitted the best automatic run; for comparison, we show the second-best runs, marked conditions (4c) and (5c), respectively.

Within each round, our submissions provided limited constrastive and ablation runs that highlight different aspects of the Expando-Mono-Duo pattern.
Since each team was only allowed to submit a limited number of runs (three in rounds 1 through 4 and eight in round 5), it was not possible to examine all interesting conditions.
Also, due to the rapid pace of the evaluation, we were only able to ``roll out'' components in our Expando-Mono-Duo design pattern incrementally.
We discuss results from each round below:

\smallskip \noindent {\bf Round 1.}
Our monoT5-3B reranker was introduced in round 1, shown as condition (1b), although its effectiveness was still quite far behind the state of the art.
The best submission in round 1 was a fusion-based run using the the vector space model, involving no neural techniques at all.
A post-hoc analysis diagnosed the issue with our run to be with first-stage retrieval:\ we had used the paragraph index, but simply replacing our pipeline with the abstract index improved nDCG@10 from 0.5223 to 0.5702.
It is unclear whether this difference was due to an evaluation artifact.

\smallskip \noindent {\bf Round 2.}
Starting in round 2 for the remaining rounds, the \run{fusion1} and \run{fusion2} baselines described in Section~\ref{settings:trec-covid} were made available to the community to serve as first-stage retrieval.
We took advantage of these strong keyword baselines as the input to our reranking pipeline.
In condition (2b), we reranked \run{fusion1} and \run{fusion2} with monoT5-3B, and then combined the reranked results using reciprocal rank fusion~\cite{cormack2009reciprocal}.
This yielded effectiveness that was, for all practical purposes, the state of the art (behind the top scoring run by only 0.001).

\smallskip \noindent {\bf Round 3.}
As the rounds progressed in TREC-COVID, we introduced more components in our Expando-Mono-Duo design.
Our duoT5-3B reranker was introduced in round 3, shown as condition (3b):\ similar to the round 2 configuration, we reranked \run{fusion1} and \run{fusion2} with the two-stage monoT5/duoT5 pipeline, and then combined the reranked results using reciprocal rank fusion.
Condition (3c) represents the same technique as condition as (2b), and thus conditions (3b) and (3c) isolate the effects of the duoT5 reranker.
We see that duoT5 contributes to small increases in nDCG@10 and average precision, but a small decrease in precision at rank 5.

The highest scoring automatic run submitted in round 3 was \run{SFDC-fus12-enc23-tf3}, which is 2.4 points higher than our best run; condition (3b) represented the second-highest automatic run.
Thus, it would be fair to characterize the monoT5/duoT5 pipeline as close to the state of the art for this round.

\smallskip \noindent {\bf Round 4.}
Our doc2query-T5 document expansion technique was introduced in round 3, thereby completing our Expando-Mono-Duo design pattern.
Row (4b) represents the same technique as row (3b), and thus conditions (4a) and (4b) isolate the effects of document expansion.
We see that document expansion clearly contributes to a large increase in terms nDCG@20 and precision at rank 20, although, interestingly, average precision decreased.

\smallskip \noindent {\bf Round 5.}
Since we were allowed eight submissions in this final round, we submitted two more ablations on our design pattern.
Row (5a), (5b), and (5e) represent the same techniques as row (4a), (4b), and (3c) respectively.
Row (5d) added the new condition where we just ran the ``Expando-Mono'' part of our design pattern.
Comparing conditions (5a) and (5d) vs.\ (5b) and (5e), we see the effects of document expansion in isolation (i.e., fixing the other parts of the pattern).
Consistent with results from the other retrieval tasks, we see large improvements in effectiveness.
Comparing (5a) and (5b) vs.~(5b) and (5e), we see that pairwise reranking with duoT5-3B contributes effectiveness gains over just reranking with monoT5.
Again, this is consistent with results from other tasks.

\section{Summary and Conclusion}

In this paper, we propose Expando-Mono-Duo T5, a design pattern for multi-stage ranking architectures that combines and synthesizes our previous work in document expansion, pointwise ranking, and pairwise ranking.
Our findings can be summarized as follows:

\begin{itemize}

\item It is possible to adapt sequence-to-sequence models for document expansion (``Expando''), pointwise ranking (``Mono''), and pairwise ranking (``Duo'').
Document expansion is a straightforward sequence-to-sequence transformation, while reranking can be accomplished by designing appropriate input sequence templates and probing model outputs to extract a probability of relevance.

\item Document expansion (``Expando'') is an effective way to improve retrieval effectiveness without requiring computationally expensive neural inference at query time.

\item Pointwise reranking (i.e., relevance classification) with monoT5 is highly effective with different first-stage retrieval results (with or without document expansion, with or without pseudo-relevance feedback).

\item Pairwise reranking with duoT5 further improves the effectiveness of monoT5 output, particular in early-precision metrics.

\item Overall, the benefits from all three components of our Expando-Mono-Duo design pattern are additive and cumulative.

\end{itemize}

\noindent These findings are supported by experimental results from the MS MARCO passage and document ranking tasks, the TREC 2020 Deep Learning Track, and the TREC-COVID challenge.
In absolute terms, we achieve results on these ranking tasks that are close to or at the state of the art.
Overall, our Expando-Mono-Duo design pattern provides a foundation and reference for transformer-based multi-stage ranking architectures.

\section{Acknowledgments}

This research was supported in part by the Canada First Research Excellence Fund and the Natural Sciences and Engineering Research Council (NSERC) of Canada.
In addition, thanks to Google Cloud for credits to support some of our experimental runs.
Finally, we would like to thank Ruizhou Xu and Hang Cui for their help in preparing the TREC 2020 Deep Learning Track runs.

\bibliographystyle{ACM-Reference-Format}
\bibliography{main}

\end{document}